# Energy transfer and interference by collective electromagnetic coupling

Mayte Gómez-Castaño[1,2], Andrés Redondo-Cubero[3], Lionel Buisson[1], Jose Luis Pau[3], Agustin Mihi[2], Serge Ravaine[1], Renaud A. L. Vallée[1], Abraham Nitzan[4,5] and Maxim Sukharev[6,7]

[1] CNRS, University of Bordeaux, CRPP, UMR5031, 115 av. Schweitzer, 33600, Pessac, France.

[2] Institut de Ciència de Materials de Barcelona (ICMAB-CSIC), Campus de la UAB, 08193 Barcelona, Spain.

[3] Electronics and Semiconductor Group, Department of Applied Physics, Universidad Autónoma de Madrid, E-28049 Madrid, Spain.

[4] Department of Chemistry, University of Pennsylvania, Philadelphia, PA 19104, USA.

[5] School of Chemistry, Tel Aviv University, Tel Aviv, Israel

[6] Department of Physics, Arizona State University, Tempe, Arizona 85287, USA.

[7] College of Integrative Sciences and Arts, Arizona State University, Mesa, Arizona 85201, USA.

**Abstract**: The physics of collective optical response of molecular assemblies, pioneered by Dicke in 1954, has long been at the center of theoretical and experimental scrutiny. The influence of the environment on such phenomena is also of great interest due to various important applications in e.g. energy conversion devices. In this manuscript we demonstrate both experimentally and theoretically the spatial modulations of the collective decay rates of molecules placed in proximity to a metal interface. We show in a very simple framework how the cooperative optical response can be analyzed in terms of intermolecular correlations causing interference between the response of different molecules and the polarization induced on a nearby metallic boundary and predict similar collective interference phenomena in excitation energy transfer between molecular aggregates.

Collective optical response of atomic and molecular emitters has been discussed throughout the evolving science of light-matter interaction. In particular, Dicke theory of superradiance emission[1] and variants such as superfluorescence[2] have been extensively discussed and different aspects of the phenomenon, including its quantum and classical aspects[3] as well as the effect of static disorder (inhomogeneous broadening) and dephasing (homogeneous broadening) were elucidated.[3-7] More recent discussions of cooperative molecular effects have focused on strong coupling between molecular excitons and plasmons or cavity modes, mostly manifested through the $\sqrt{N}$ scaling of the observed Rabi splitting in such systems, where N is the number of involved molecules.[8-10] It should be noted that while both phenomena are manifestations of collective response, Dicke superradiance is observed during the decay of a state in which all molecules in a given cluster are initially excited while the collective Rabi splitting usually involves the bright single exciton state where a single molecule excitation is delocalized over N molecules. Significantly, the radiative lifetime of such a bright exciton, which carries most of the oscillator strength of the *N*-molecule cluster, is $\tau_R(N) = \tau_R(1)/N$, a property which expresses itself also in energy transfer out of such cluster. [11] Both this collective feature and Dicke's superradiance physics have been discussed recently with regards to their implications for the performance of energy conversion devices. [12-21] It should be noted that the theme exploited in all these studies is the correlated behavior of a many-body system (often modeled as a cluster of *N* two-level atoms) supported by their mutual coupling to the radiation field, sometimes in the form of strong coupling to a cavity mode. The general physics of such systems has been extensively discussed since the middle of the previous century. [22-31] More recent studies have placed such systems in the vicinity of metal nanostructures, where interaction between molecular emitters is mediated not only by free or cavity photon modes but also via their mutual interactions with metal plasmons. [11, 32-38]

While not usually phrased in this language, cooperative optical response can often be analyzed as interference between the response of different molecules, where correlations determine whether one observes constructive or destructive interference. In the simplest case of a single excited molecule placed in front of a mirror, emission from the molecule and the polarization induced in the metal (or the molecule and its image) interfere, leading to oscillations in fluorescence lifetimes as a function of the distance from the surface.[39-43] Obviously, this phenomenon is not limited to a single molecule. The response of a molecular cluster, including all the intricacies mentioned above, is expected to interfere with the response of the corresponding polarization induced on a nearby metallic boundary or a metal particle. Indeed, such effects have been very recently discussed[37] and observed. [44]

In this paper we experimentally and theoretically demonstrate such collective molecular phenomena. We note that with strong dependence on intermolecular correlations, specific behaviors may reflect properties of specific structures. By studying the optical properties of such structures, we can characterize their collective nature as reflected in their steady state and relaxation behaviors, their dependence on emitter density (or number), system dimensionality and structure as well as effects of structural and dynamical disorder. Our experimental results demonstrating such collective behavior are presented in Section 2. We discuss these phenomena within a simple theoretical model in Section 3, and present results of numerical simulations based on a Maxwell-Bloch model for the radiation field–molecules system. Even on this simple level we can demonstrate dramatic effects of structure, order and emitter density, implying new ways to control the optical properties of molecular nanodevices.

**Experimental results.** The proposed samples are fabricated as described in Fig. 1a (see Methods for fabrication details). First, a flat gold substrate is covered by a silica layer with a well-controlled thickness. Such film works as a spacer between the metallic mirror and a post deposited molecular layer of Atto 655 fluorophores, whose number density can be varied with the concentration of molecules in solution. The fluorescent dye was chosen due to its characteristic features of strong absorption (1.25 $10^5$ M-1 cm-1), high fluorescence quantum yield (0.3) and high thermal and photo-stability for a red dye emitting at 680 nm, e.g. at a wavelength where autofluorescence of the sample is drastically reduced, ensuring an extremely clean collection of the photon counts. A layer of polymethyl methacrylate (PMMA) is finally deposited on top of the structure in order to fix the position of the fluorescent dyes and provide similar electromagnetic boundary conditions on both sides of the layer. As a reference sample, the ensemble of fluorophores and PMMA were deposited in the same conditions on a bare glass substrate. Once prepared, the samples underwent structural and ellipsometry characterizations, ensuring that they were designed as prescribed. We then investigated the samples by time resolved fluorescence microscopy in order to determine the emission properties of the system. As presented in Fig. 1b, a pulsed laser operating at 654 nm is normally focused on the sample surface. The emitted fluorescence is then collected through the same objective and sent to the detector for the following reconstruction of the decay profiles.

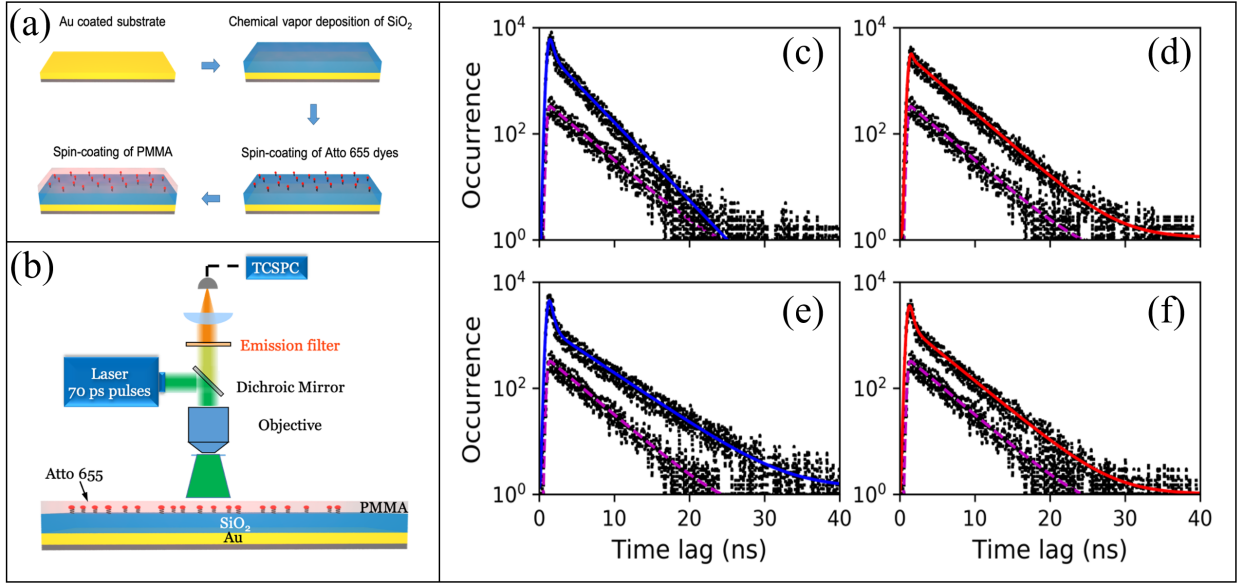

**Figure 1**. Panel (a) shows the scheme of the fabrication process. Panel (b) shows the experimental setup for optical measurements. Panels (c)-(f) show experimental fits of the decay profiles of ensembles of molecules deposited either on a bare glass substrate (black dots fitted by a magenta line) or at a given distance, fixed by a silica layer, on top of the gold mirror structure. In the former case, the decay is mono-exponential with a rate $\gamma$ = 0.178 µeV for a molecular layer's density estimated at $3 \times 10^{22}$ m$^{-3}$. In the latter case, the decay is found to be best fitted by a bi-exponential with rates varying with both the molecular density and the distance to the gold mirror. The respective values for $\gamma_L$ ($\gamma_S$) are (c) 1.68 (0.224) µeV for the $3 \times 10^{22}$ m$^{-3}$ layer density at a 80 nm distance (blue line); (d) 5.52 (0.191) µeV for a $3 \times 10^{23}$ m$^{-3}$ layer density at a 380 nm distance (red line); (e) 1.64 (0.145) µeV for a $3 \times 10^{22}$ m$^{-3}$ layer density at a 200 nm distance (blue line); (f) 1.90 (0.184) µeV for a $3 \times 10^{23}$ m$^{-3}$ layer density at a 560 nm distance (red line) from the gold mirror structure.

The decay profiles of Atto655 molecules deposited on a bare glass substrate from a $10^{-8}$ M concentration of dyes in ethanol exhibit a clear single exponential decay profile with a decay rate of 0.178 µeV (Fig. 1c-1f: black dots fitted by a magenta dashed line, shown in each panel as a reference). The decay profiles of the same molecules deposited at a controlled distance (silica spacer) of the gold mirror structure significantly deviate from a single exponential decay profile. A bi-exponential function $I(t) = I_S \exp(-\gamma_S t) + I_L \exp(-\gamma_L t)$ convoluted to the instrumental response function (IRF) of the setup provides a good estimate of the small $\gamma_S$ and large $\gamma_L$ decay rates (Fig. 1c-1f). Both rates depend on the molecular layer – gold mirror distance as well as on the molecular density of the film. We show results for molecular number densities $3\times10^{22}$ m$^{-3}$ (Fig. 1c and 1e) and $3\times10^{23}$ m$^{-3}$ (Fig. 1d and 1f) and for molecular layer to the gold mirror distances in the performed range extending from 20 nm up to 680 nm by step of 60 nm.

One notices immediately the shortening of the fast decay, $\gamma_L$, while going from lower (Fig. 1c and 1e) to higher (Fig. 1d and 1f) concentrations of molecules, a feature not observed for the slow decay $\gamma_S$. This rapid behavior, as better exhibited in the fitted values reported in Fig. 2a and 2b, indicates a collective relaxation behavior. Experiments were also performed at even larger concentration (up to $3\times10^{25}$ m$^{-3}$), however the fast decay rate was beyond the time resolution of our apparatus: we could then only see a single exponential profile and were able to retrieve the small decay rate, $\gamma_S$.

Figure 2 exhibits the evolution of these decay rates as functions of number density and distance D from the gold mirror structure. Obviously, the long decay rates exhibit damped oscillations (Fig. 2c and 2d), as already observed in Drexhage's experiments over 50 years ago [39]. These observations were shown to result from interference between an emitter and its image in the mirror.[39, 40, 42, 43, 45]. Strikingly enough, this behavior is not limited to individual molecules but is also seen in the collective relaxation. Remarkably, the $\gamma_L$ oscillations are not damped over the observed distance, as seen by comparing Figs. 2a and 2c. It should be noted that the fast decays exhibited in Fig. 2b are on the edge of our time resolution. The error bars shown in this figure are merely standard deviations of various measurements performed on the same sample and do not reflect the setup's time resolution. As such, not all decay profiles could be fitted appropriately. Nevertheless, one can clearly see that the large decay rates are significantly higher for the $3\times10^{23}$ m$^{-3}$ layer density (Fig. 2b) than for the $3\times10^{22}$ m$^{-3}$ layer density (Fig. 2a). For the largest concentrations of molecules (up to $3\times10^{25}$ m$^{-3}$), the trend followed by the small decay rates is similar to those shown in Figs. 2c and 2d.

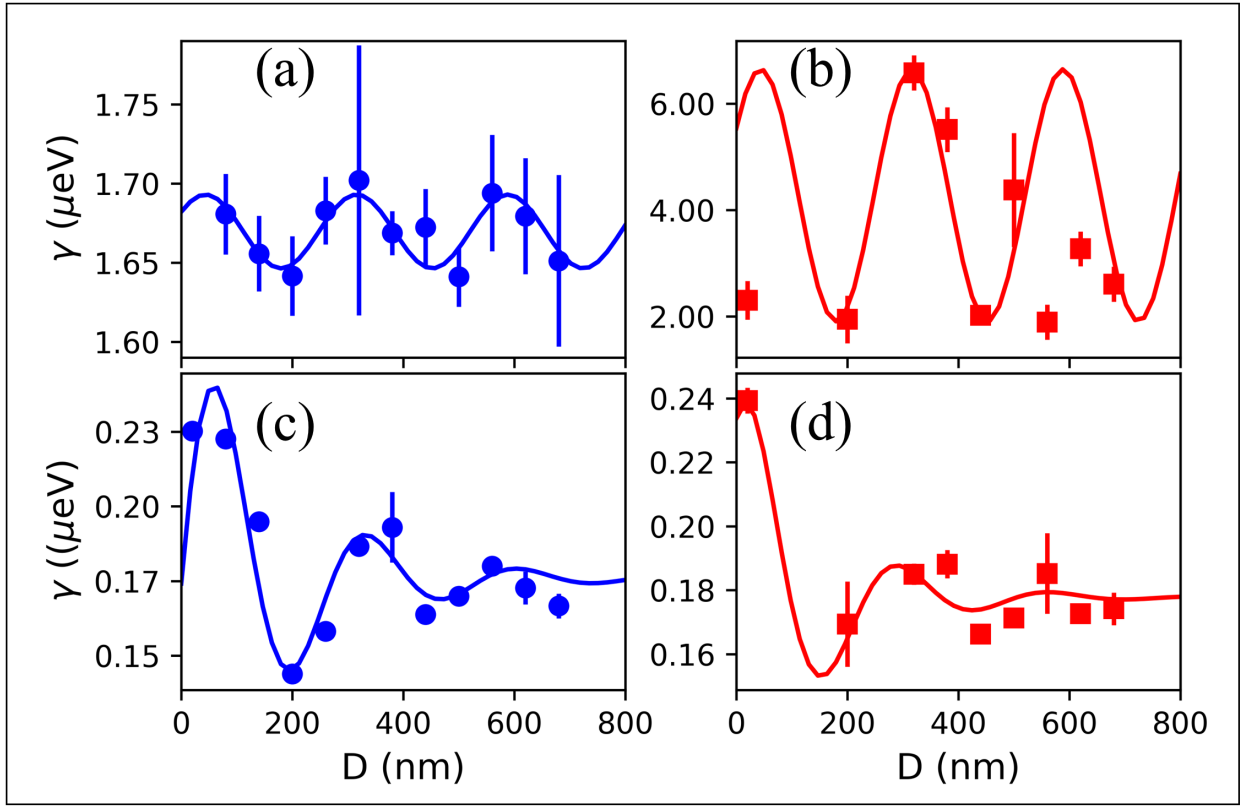

*Figure 2*. Evolution of the small and large decay rates as a function of the molecular layer density and distance from the gold mirror structure. Panels (a, b respectively) and (c, d respectively) show the evolution of $\gamma_S$ and $\gamma_L$ for a $3\times10^{22}$ m$^{-3}$ ($3\times10^{23}$ m$^{-3}$ respectively) layer density. Experimental $\gamma_L$ (dots and squares) in (a) and (b) were fitted by an appropriate sinusoidal signal while the $\gamma_S$ (dots and squares) in (c) and (d) were fitted by a damped sinusoid. The frequency (related to the emission frequency of the molecules and the refractive index of the surrounding medium) found in a) was used as a fixed parameter for the three other fits. The amplitude of the sinusoid in (b) was merely adjusted by hand, which makes it more a guide to the eye than a fit.

**Theoretical model and computational results.** For the simplest model of collective response from a molecular layer, we assume the molecules are oriented parallel to the gold mirror thus leading to a one-dimensional electromagnetic setup. The corresponding Maxwell's equations governing the electrodynamics of molecule-to-molecule interactions are

$$\frac{\partial B_y}{\partial t} = -\frac{\partial E_x}{\partial z}, \qquad (1)$$
$$\varepsilon_0 \frac{\partial E_x}{\partial t} = -\frac{1}{\mu_0}\frac{\partial B_y}{\partial z} - J_x,$$

where *z* corresponds to the longitudinal coordinate perpendicular to the mirror, $J_x$ is the polarization current density of the molecular layer. The molecules are assumed to be two-level emitters described by the density matrix $\hat{\rho}$ and driven by a local field in accordance with

$$i\frac{d}{dt}\begin{pmatrix} \rho_{11} & \rho_{12} \\ \rho_{21} & \rho_{22} \end{pmatrix} = \begin{pmatrix} \frac{E_x \mu_{12}}{\hbar}(\rho_{12} - \rho_{12}^*) & -\frac{E_x \mu_{12}}{\hbar}(\rho_{22} - \rho_{11}) - \omega_{12}\rho_{12} \\ -\frac{E_x \mu_{12}}{\hbar}(\rho_{22} - \rho_{11}) - \omega_{12}\rho_{12}^* & \frac{E_x \mu_{12}}{\hbar}(\rho_{12}^* - \rho_{12}) \end{pmatrix}. \quad (2)$$

Here $\mu_{12}$ and $\omega_{12}$ are the transition dipole and the transition frequency of a single molecule, respectively. The polarization current is then calculated as

$$J_x = \frac{\partial P_x}{\partial t} = n_0 \mu_{12} \frac{\partial}{\partial t}(\rho_{12} - \rho_{12}^*), \quad (3)$$

where $n_0$ is the number density of molecules.

We consider the dynamics of molecules initially prepared by a sudden (short time) excitation of all molecules in a coherent superposition state such that the population of excited molecules is significantly less than 1 throughout the time propagation (this corresponds to the optical linear regime). We set it to 0.01, although we note that for the range of molecular densities and material parameters considered in this manuscript an excited state population as high as 0.5 lead to qualitatively similar results. Thus, the initial conditions for the coupled Maxwell-Bloch equations are

$$\begin{aligned} \rho_{11} &= 1 - \rho_{22} = 0.99, \\ \rho_{12} &= \sqrt{\rho_{11}\rho_{22}} \exp(i\varphi), \end{aligned} \quad (4)$$

where the initial phase $\varphi$ can be taken the same (and then chosen to be 0) for all molecules or with a random component to represent disorder, as discussed below.

The dynamics of molecules obtained from Maxwell-Bloch equations with the initial conditions (4) shows that the ensemble average excited state population as a function of time follows an exponential decay law, exp(-$\gamma t$). The rates $\gamma$, calculated by fitting the numerical time dependent excited state population, are displayed in Fig.3.

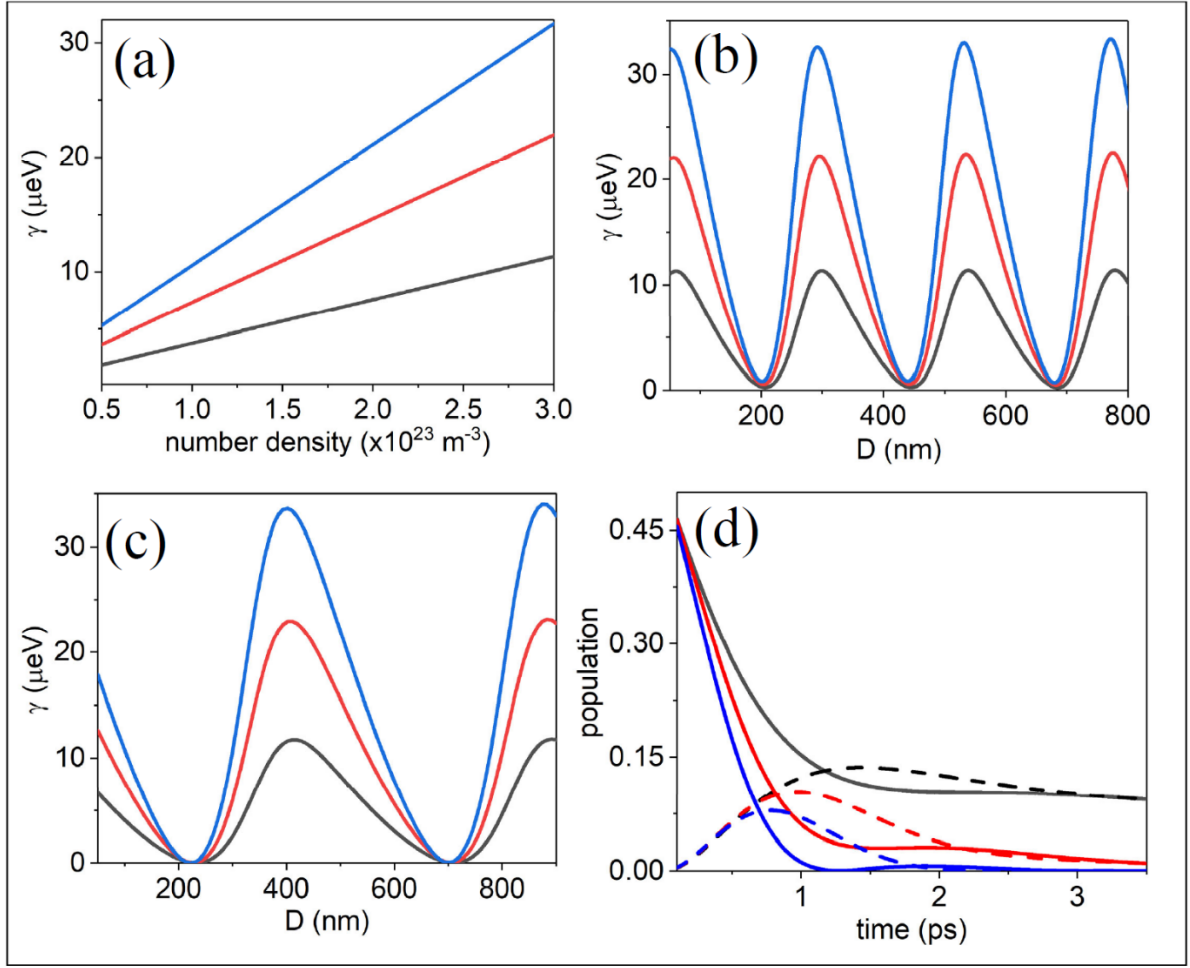

*Figure 3*. Panels (a) and (b) show simulations for a single molecular layer in the experimental geometry (Fig. 1a and 1b). Panel (a) shows the decay rate extracted from the ensemble average excited state population as a function of the molecular density for three different thicknesses of the molecular layer: black – 5 nm, red – 10 nm, and blue – 15 nm. The thickness of the silica spacer is 300 nm. Panel (b) shows the spatial dependence of the decay rate on the thickness of the silica layer (distance from the gold mirror) for the same three molecular layer's thicknesses as in panel (a). The molecular number density is $3\times10^{23}$ m$^{-3}$. Panels (c) and (d) explore the collective evolution in a system of two molecular layers separated by a silica spacer without mirrors. In (c) both layers start in the same coherent superposition as defined by eq. (4) and the decay of the overall excited state population is shown as a function of the interlayer distance (thickness of a silica spacer) for the same three molecular layer's thicknesses as in panel (a), with the molecular number density in both layers taken as $3\times10^{23}$ m$^{-3}$. The initial state in (d) corresponds to one of the layers prepared in the state (4) while the molecules in the other layer are initially in the ground state, and the dynamics of interlayer energy transfer and overall relaxation is observed. Black lines correspond to spacer thickness of 20 nm, red lines are for 50 nm, and blue lines

correspond to 80 nm. Other parameters are: the molecular number density is $10^{26}$ m$^{-3}$, the layers' thickness is 10 nm. In all simulations the phase of the coherence in Eq. (4) is set to 0.

Fig. 3a shows the dependence of the decay rate on the molecular number density for different thicknesses of the molecular layer according to the experimental geometry (Fig. 1a and 1b). It is seen that the decay rate scales linearly with the molecular number density, as expected from a delocalized single exciton state [33]. For a silica thickness of 300 nm and a molecular number density of $3\times10^{23}$ m$^{-3}$ our model calculation predicts the fast (collective) decay rate of 22 μeV, which overestimates the experimental fast rate of 6.6 μeV (Fig. 2b). The overestimate is likely due to the fact that the calculation takes all molecules to be perfectly aligned parallel to the interface and to each other. In contrast, for an isotropic molecular system the averaging over all orientations of the molecular dipole leads to an effective coupling between the macroscopic polarization and the electric field which is reduced by the factor of 1/3 [9]. This results in a decay rate of 7.3 μeV, remarkably close to the observed rate.

Next consider the results presented in Fig. 3b, which shows the observed decay rate as a function of the distance between the molecular layer and the gold film (as per experimental setup this corresponds to varying the thickness of the silica layer). Since the light emitted by the molecules is reflected by the mirror it undergoes a π-phase change (for an ideal mirror). Thus, the resonant condition corresponding to the destructive interference of the reflected and emitted waves leading to a near-zero decay rate is

$$n_{min} \cdot \lambda_{12} = c \cdot t_{round\ trip} = c \cdot \frac{2 \cdot d}{c} \Rightarrow d = n_{min} \cdot \lambda_{12}/2, n_{min} = 1, 2 \ldots \quad (5)$$

where $c$ is the speed of light in the media between the mirror and molecules, $d$ is the layer-to-mirror distance, and $\lambda_{12}$ is the molecular transition wavelength (since molecules are embedded in PMMA the corresponding wavelength is shorter by the factor of the PMMA refractive index compared to its value in vacuum). Small deviations from (5) are attributed to the finite thickness of the molecular layer (as clearly seen it affects the positions of minima in Fig. 3b) and the phase change of the emitted radiation when it bounces off of a gold surface. There is also a small mismatch of the refractive indices for the silica and PMMA, in which the molecules are embedded. Nonetheless the spatial modulations of the decay rate obtained from one dimensional Maxwell-Bloch equations agree perfectly with experimental results. Importantly, this behavior, long known for individual molecules, is predicted and observed also for the collective decay of a molecular aggregate, here a molecular layer parallel to the mirror, as indicated by the density dependence of the rate seen in Fig. 3a.

It is also important to note that dephasing of intermolecular coherence is what causes the transition from collective (fast decay) to single molecule dynamics (slow decay) observed experimentally. In the present one-dimensional model such dephasing, associated with dynamic or static disorder (inhomogeneous broadening), can be imposed only partially along the direction normal to the layers, since the molecular layers respond collectively by the nature of the 1-d calculation. Thus, in the simulations discussed in Fig. 3 the phase in (4) is taken the same for all system units, i.e. all points of our one-dimensional grid. If the phase is chosen randomly between 0 and $2\pi$ for each grid point, representing system disorder, the same oscillations in the decay rate vs. distance are seen, albeit with a significantly reduced amplitude than experimentally observed. Another important difference between the behaviors of the coherent and incoherent molecular systems lies in the observed distance dependence. For the geometry studied here, a system with a fully coherent molecular layer is truly one-dimensional, implying that the rate oscillates with a distance independent amplitude, as seen in the experimental fast decay (Fig. 1c-1f) and captured by our one-dimensional simulations. In contrast, when the emitters oscillate with random phases, the observed emission is a sum of signals generated by individual emitters so that the pertinent geometry is three-dimensional, characterized by an oscillation amplitude that decays with distance as already reported in Refs. [39, 40] and seen in Fig. 2c, 2d. The presence of an incoherent component in the molecular response may also explain why a perfect destructive interference is not observed experimentally.

Our successful modeling of the experimental observation implies the important prediction that collective interference phenomena should be observed for interacting molecular assemblies. This is demonstrated in Figs. 3c and 3d where the two assemblies are taken as two parallel molecular layers separated by a silica spacer. One can envision various scenarios for the time dynamics depending on initial conditions. If molecules in both layers are prepared in the same coherent superposition (Fig. 3c), we observe the familiar exponential decay as in Fig. 3a and 3b depending on spacer's thickness, albeit with a change of phase: for a "real" molecular layer (as opposed to an "image" layer) the condition for the destructive interference between electromagnetic waves emitted by both layers is $d = (n_{min} + 1/2)\lambda_{12}$, $n_{min} = 0,1,...$, showing a phase shift compared to Eq. (5), as seen in Fig. 3c. Again, the resonant condition is clearly maintained throughout all spacer's distances with a slight deviation due to the finite molecular layers' thickness.

In a different scenario (Fig. 3d), only molecules in one of the layers are prepared in the excited coherent superposition. Here, collective relaxation and energy transfer are seen in the time evolution. The time dependence of the excited state populations reveals complex dynamic exhibiting a clear sign of interference as the decay rate is significantly influenced by the thickness of the silica spacer. While the excited layer begins to decay emitting the radiation, the initially relaxed molecules are picking up this radiation. The time at which the initially relaxed molecular layer reaches its maximum excitation clearly depends on the molecular density and spacer's thickness, a clear predicted evidence of the collective nature of energy transfer.

**Conclusion.** In summary, we demonstrated both experimentally and theoretically that the collective decay rates of molecular assemblies strongly depend on the distance between metal and molecules. They exhibit spatial modulations that are attributed to constructive/destructive interferences. Experimentally, two distinct decay rates are observed: the fast decay oscillates with the distance from mirror without spatial damping whereas the slow decay exhibits similar oscillations but showing a significant spatial damping. The latter is due to single molecule fluorescence[39, 40, 41, 42, 43] while the former is of collective nature. Simple theoretical model based on one-dimensional Maxwell-Bloch equations agrees well with experimental characteristics of the observed fast decay and provides guidance to future experiments aiming to control energy transfer at the nanoscale.

**Acknowledgements.** This work is sponsored in part by the Air Force Office of Scientific Research under Grants No. FA9550-15-1-0189 and FA9550-19-1-0009 (M.S.), the U.S. National Science Foundation, Grant No. CHE1665291 (A.N.), CTQ2017-84309-C2-2-R (ARC and JLP), P2018/NMT4349 Comunidad de Madrid (ARC and JLP), RYC-2015-18047 (ARC), and the US-Israel Binational Science Foundation, Grant No. 2014113 (M.S. and A.N.). We thank the Spanish Ministry of Economy and Competitiveness (MINECO) for its support through Grants MAT2016-79053-P and grant SEV-2015-0496. This project has received funding from the European Research Council (grant agreement No 637116). This work was also financially supported by the French National Research Agency (ANR) as part of the Initiative for Excellence IdEx Bordeaux program (ANR-10-IDEX-03-02).

# Supplemental material: energy transfer and interference by collective electromagnetic coupling

Mayte Gómez-Castaño[1,2], Andrés Redondo-Cubero[3], Lionel Buisson[1], Jose Luis Pau[3], Agustin Mihi[2], Serge Ravaine[1], Renaud A. L. Vallée[1], Abraham Nitzan[4,5] and Maxim Sukharev[6,7]

[1] CNRS, University of Bordeaux, CRPP, UMR5031, 115 av. Schweitzer, 33600, Pessac, France.

[2] Institut de Ciència de Materials de Barcelona (ICMAB-CSIC), Campus de la UAB, 08193 Barcelona, Spain.

[3] Electronics and Semiconductor Group, Department of Applied Physics, Universidad Autónoma de Madrid, E-28049 Madrid, Spain.

[4] Department of Chemistry, University of Pennsylvania, Philadelphia, PA 19104, USA.

[5] School of Chemistry, Tel Aviv University, Tel Aviv, Israel

[6] Department of Physics, Arizona State University, Tempe, Arizona 85287, USA.

[7] College of Integrative Sciences and Arts, Arizona State University, Mesa, Arizona 85201, USA.

*Experimental methods.* The samples were designed as shown in Figure 1a. Commercial optically opaque gold films of 150 nm thickness grown on quartz slides (ACM, France) were taken as substrates for this study. Dielectric layers of silica ($SiO_2$) were deposited on top of these substrates by electron cyclotron resonance plasma enhanced chemical vapour deposition (ECR-PVD) using a PlasmaQuest 357 equipment. The microwave power was set at 1000 W, working with 60 sccm of $SiH_4$ (diluted at 5% in Ar) and 70 sccm of $O_2$. Depositions were carried out at room temperature at a working pressure of $10^{-6}$ mbar. The layers thickness was varied by the deposition time: the growth rate was determined by spectroscopic ellipsometry on Si reference samples to be 17 nm/min. Atto 655 fluorophores (Sigma-Aldrich, 634 g/mol) were diluted in ethanol at different concentrations ($10^{-8}$ M, $10^{-7}$ M, $10^{-6}$ M and $10^{-5}$ M) and spin-coated using a SCS G3-8 system. A fixed volume of 20 μL was poured onto the samples and spun at 3000 rpm for 20 s, being the acceleration and deceleration times of 3 s. The molecular number density of the layers was estimated to be $3 \times 10^{23}$ m$^{-3}$ and $3 \times 10^{22}$ m$^{-3}$, for the lowest concentrations. Polymethyl methacrylate (PMMA) powder purchased from Sigma-Aldrich (Mw ~ 15000) was diluted in acetone at 1:80 weight ratio and spin-coated on top of the samples to ensure the fixed position of the fluorophores.

Aliquots of 40 µL of this solution were spun at 3000 rpm for 30 s, with 0.6 s of acceleration and deceleration times, ensuring a 50 nm thickness of the final PMMA layer.

We built the custom-made setup shown in Figure 1b. The excitation light is provided by 70 ps width pulses delivered by an EPL-655 (Edinburgh Instrument) operating at a 654 nm wavelength, at a controlled repetition rate varied, following our needs, from 1 kHz to 1 MHz while keeping the typical peak power of 120 mW. Such pulses are then directed, owing to a dichroic mirror, and focused on the sample through a high numerical aperture objective (Zeiss Neofluar EC Epiplan - 100x/0.9DIC (422392-9900)). We did not focus too tightly, e.g. we did not fill the back of the objective in order to get a diffraction limited spot on the sample, in order to avoid extremely large incident angles, which were not included in the simulations. The emitted signal is collected through the same objective and sent to the detector (ID Quantique - ID100 single detection module) through two emission filters (Omega : 710QM80, Semrock : 664nm RazorEdge ultrasteep long-pass edge filter LP02-664RU-25). The detector is equipped with a Time Correlated Single Photon Counting Card (ID Quantique - ID800) in order to register the photons 1 by 1, in order to reconstruct the decay profiles afterwards.

*Computational details*. Molecules assumed to be two-level emitters form a uniform layer finite in one direction and infinite in two others reducing electrodynamical simulations to one dimension. The transition dipole moment $\mu_{12} = 8.5$ Debye is estimated from the experimental absorption spectra for Atto655 molecules. The system of Maxwell-Bloch equations **Error! Reference source not found.-Error! Reference source not found.** is numerically integrated using finite-difference time-domain approach [46]. When simulating open boundaries, we terminate the simulation domain by the perfectly matched absorbing layers [47]. Using a spatial resolution of 0.2 nm and a time step of $3.37 \times 10^{-4}$ fs has made it possible to achieve numerical convergence for number densities up to $2 \times 10^{26}$ molecules/m$^3$. In all simulations the transition frequency is set at 1.77 eV (700 nm). The transition dipole is 8.5 Debye. The optical response of gold is simulated using the Drude-Lorentz model with parameters taken from [48]. The refractive index for silica is 1.45. The PMMA layer's thickness is 50 nm and its refractive index is 1.53. It is also assumed that the molecules are fully covered by PMMA.